# Using a Cloud Based Collaboration Technology in a Systems Analysis and Design Course

Emre Erturk
Eastern Institute of Technology, Napier, New Zealand

*Abstract*— In order to effectively prepare the next generation of IT professionals and systems analysts, it is important to incorporate cloud based online collaboration tools into the coursework for developing the students' cooperative skills as well as for storing and sharing content. For these pedagogical and practical reasons, Google Drive has been used at a medium-sized institution of higher education in New Zealand during the Systems Analysis and Design course. Ongoing and successful use of any learning technology requires gathering meaningful feedback from students, and acting as a mentor during their learning journey. This study has been developed and implemented to help students enjoy the collaborative technology and to help increase their satisfaction and commitment. In order to overcome the obstacles that may prevent students from using Google Drive optimally, an initial survey has been conducted to better understand the influential factors and issues. Furthermore, this study aims at promoting various types of collaboration and sharing: seeing and learning from other students' work, receiving direct suggestions from others, and allowing others to edit documents that belong to them. Following the results of the first quantitative survey, numerous teaching strategies were formulated and implemented. A final qualitative survey was done at the end of the course for students to evaluate their project work. The results of this study also provide original practical and theoretical implications that may be of interest to other researchers, course designers, and teachers.

## I. Introduction

The Systems Analysis and Design (SAD) course is an important part of the information technology (IT) curriculum for many institutions of higher education around the world. This study was conducted at a medium-sized institution of higher education in New Zealand, which delivers a face-to-face Systems Analysis and Design course to students on its main campus, including a second section running via blended delivery for a cohort at its remote campus. This course expects a significant amount of self-directed hours from students for reading and working on the assignments, in addition to lectures and tutorials. The lectures are delivered via video conference sessions to the remote campus. This study encompasses the whole course since the shared online course site and online applications play a great role for both campuses, for providing electronic resources to all students and facilitating additional learning activities.

SAD helps prepare students for job roles such as IT project manager, business analyst, and systems developer. Employment of business and systems analysts is growing. Industry stakeholders expect to receive future graduates who can adapt quickly to the workplace by virtue of practical and interpersonal skills gained during their study. Business and systems analysts work on systems and software development projects and prepare high quality documentation and prototypes. They collaborate and communicate with a variety of stakeholders including other analysts, programmers, non-technical end-users, and executive sponsors.

In this course, there is a group assignment that provides a valuable interpersonal learning experience and an intensive opportunity for students to apply their newly learnt methodology, tools, and skills, such as communication and fact-finding. During this assignment, students form small teams and carry out a complex project together. The success of these teams is not a foregone conclusion, especially in a challenging project requiring both good social and IT skills.

Google Drive is offered by Google as part of its ecosystem of web based services. It provides access to data storage, a suite of office applications, synchronous editing, permissions control, and instant communications. Due to its popularity, Google Drive is a suitable and free collaboration tool for students, and can serve as a good repository for their systems development documentation. Effective group work is an essential skill for students for becoming work-ready. This technology assists them with both the technical and social aspects of their learning journey during SAD. Furthermore, in light of limited prior research on this topic, the study of how Google Drive was implemented, evaluated, and managed in an important course within an IT curriculum, provides original findings and perspectives.

## II. Literature Review

It is necessary to better understand the context of teaching systems analysis and design, particularly in terms of the following: facilitation of effective learning, the use of educational technology in a blended environment, and which approaches to take for targeting the improvement of social skills and collaboration during typical project activities.

Course learning activities are not only for transferring knowledge but also for developing students' overall learning skills and attitudes. It is also important for teachers to help refine and reinforce the knowledge that the students gain from the lectures and their own reading. In addition to the theory, the learning activities also need to help build and activate the students' personal skills and attitudes related to systems analysis and design. As with any other challenging course, the learner's journey in SAD requires active participation and self-monitoring as well as a high degree of working in groups and building knowledge together.

Blended learning involves a mixture of both online (distance) delivery and traditional face-to-face classroom instruction. SAD is a blended or sometimes a purely online course at many universities and institutions around the world. Interactivity between everyone is pivotal to online delivery. E-learning provides interesting opportunities for sharing information by the teacher as well as between the learners themselves. This shared information can continuously change and expand with the contributions of both teachers and active learners.

It is also important to understand the preferences of learners within the course's online LMS, i.e. the learning management system (such as Moodle, Blackboard, Desire2Learn, etc.). One of the most important expectations of students from an LMS's user interface is the ease of integration with other applications [5]. By taking advantage of other web based and social applications, knowledge and other artefacts can continue to be available after the course, and can be shared with a larger audience.

The categories of features offered by learning management systems and how effectively or frequently those features are used by an institution are discussed in many papers. A convenient model with five levels (Levels 0 through 4) was proposed by Janossy [8]. A similar model was also adapted by Abazi-Bexheti, Kadriu, and Ahmedi [1] with some modifications, which, for the most part, involved changing the highest level. Their highest level looks at the extent to which students are sharing knowledge and co-developing course resources. This can also be accomplished with the assistance of third party applications and learning technologies. Student involvement in creating course content can be supported by Google Drive. Examples and artefacts from the previous semesters' student projects can be shared with new students. This way, students are not only engaging in a peer teaching role while collaborating with their own team members but also contributing to the learning activities of future students.

With the increasing abundance of web and cloud based applications and open educational resources, it is no longer necessary for a learning management system to provide all desired or interesting features by itself. Online learning activities often involve collaboration and contributing to other learners' knowledge. In this context, open source software provides more freedom and flexibility to schools and users, as to ownership and customization of content.

There are many examples of open access, open source, or free software that can be utilized effectively within a course's broader e-learning environment on top of the main platform (e.g. Moodle). For example, YouTube is a user content driven and cloud based video repository with an open interface for other applications. YouTube videos are easily embedded in Moodle. Newly created instructional videos can also be made publicly available via YouTube. Similarly, Google Hangouts enable live conversations at no cost between lecturers and their online students. Mobile learning (m-learning), i.e. learning with the assistance of mobile devices, benefits tremendously from being able to communicate easily at no cost, and from being able to share information anywhere by using these open technologies.

As stated by Burns [4], among teachers of systems analysis and design, there are many areas of difference, in terms of learning approaches and material. However, as Burns [4] found out, there is almost a consensus in certain aspects. For example, most courses split their students into groups so they can collaborate on their projects. Similarly, 'people issues' is one of the important concepts covered and illustrated in the majority of SAD courses.

Easy to use and featuring a range of supporting applications, Google Drive is one of the tools that can facilitate this course. Furthermore, there have been institutions that have reported positive experiences with Google Drive (previously known as Google Docs). For example, according to Rowe, Bozalek, and Frantz [11], online collaboration using Google Drive enhanced the students' learning experience by providing a means for interaction, and an autonomous space outside of the classroom. Cloud based applications (e.g. Google Drive) are also a way of deploying mobile learning. Open educational resources have often been seen as providing greater flexibility to the end users, i.e. educators and students alike. Therefore, Google Drive is not only collaboration tool for students but also a recent example of open resources in education. The SAD course has been designed to take advantage of the pedagogical benefits of collaborative learning. Collaborative learning is fun for the students and allows them to experiment; it also simulates real world situations [10]. In the future, as professionals working on a project, they will frequently share and edit documents and artefacts together in a similar way.

### III. METHOD

Feedback is an important part of designing a course. Boud and Molloy [3] suggest the improvement of feedback in many ways, not just from the teacher to the students. For example, more feedback from the students to the teacher can be encouraged. Furthermore, exchanging more feedback between students themselves would be very beneficial.

Informal qualitative feedback on using Google Drive was collected for the first time in 2013 from Systems Analysis and Design students at this institution. The mixed nature of this initial unpublished feedback (reflecting both positive and negative experiences) triggered a further inquiry as to the factors behind the varying levels of satisfaction and success among students using this tool during their Systems Analysis and Design project. These factors include how well they understand the tool itself, their level of commitment to the tool, and the ways in which they choose to use Google Drive. The last of these factors (how they use the tool) includes strategies and choices regarding ownership, sharing, and editing [2]. This current study has taken up a new investigation with the aim of better understanding, promoting, and improving the collaborative use of Google Drive among the students.

The Systems Analysis and Design course (during Semester 1 of 2014) included a total of 36 undergraduate students (83% of whom were male). 92% of this group consisted of full-time students. The same percentage also applied to domestic New Zealand students (rather than international students). The mean age for this group was

25. The median age was 21. The students' ages ranged from 18 to 45 although young learners were predominant.

Two types of new feedback, i.e. qualitative and quantitative, have been collected from students during the systems analysis and design group project in 2014. First, a quantitative survey has been conducted in order to investigate the influence of three main factors on student satisfaction and success with Google Drive: 1) how well they understand the technology, 2) their level of commitment to the technology, and 3) how they choose to use the technology. The design of the questions and the Likert scale in the survey is partly based on Perlman's Usefulness, Satisfaction, and Ease of use (USE) questionnaire [9]. This first survey was conducted early in the course in order to diagnose the students' initial levels and attitudes, and to help provide appropriate instructions and support later during the course throughout the group project. At the end of the project, a second evaluation was done to understand the progress made, and the students' concluding thoughts. This qualitative student feedback was collected through a follow-up questionnaire, which encouraged the students to reflect on all aspects of their teamwork (including Google Drive) during the project assignment, which they completed.

There are two hypotheses to the initial quantitative and exploratory part of this study. First, the more often students use the technology and the more confident they feel about their skills (two variables related to factor 1), the more committed to the technology they are likely to be – their satisfaction level, how much they enjoy it, how effective they believe the technology is, and their likelihood of recommending it to others (four variables related to factor 2). The first hypothesis has been tested through a correlation analysis between these variables, as operationalized by the students' survey responses.

The second hypothesis is that students are not ready to use the technology in one or more of the following collaborative ways: seeing and learning from other students' work, allowing others to see their work as well as receiving suggestions, and allowing others to edit documents that belong to them (three variables related to factor 3). This hypothesis is also partly based on the previous literature suggesting that students exercise limited collaboration on wikis and tend to continue the practice of individual accountability and ownership [7].

For the project, each group was responsible to complete a list of deliverables. They were recommended to spread the workload by allocating some of these tasks to individuals within a group and, in turn, individual members were encouraged to cooperate with others. The advanced form of collaboration (i.e. the third variable) is where students empower and help one another – by making and allowing direct improvements, corrections, and contributions to each other's documents. However, psychological ownership with controlling or protective attitudes (on one extreme) or indifference (on the other extreme) may inhibit students from doing these. By the way, it might be useful to note that document editing on Google Drive keeps track of the revision history, showing when changes were made and by whom, and makes it possible to undo changes and revert to an older version. In order to explore the students' present attitudes toward collaboration, the second hypothesis was put forward and tested through descriptive statistics, which were calculated for the three variables in question, using the students' Likert scale survey responses.

## IV. RESULTS AND DISCUSSION

Table 1 summarizes the results of the correlation analysis for the first hypothesis. Google Drive is referred to as GD in the table. Frequency (how often the students use Google Drive) has a positive but a small correlation with the four variables associated with their commitment to Google Drive. In comparison, Skill has a much higher correlation with the same variables across the four columns in the table below. The better a student's skill at using Google Drive (i.e. in terms of his/her own perception) the greater his/her commitment and satisfaction will be, and vice versa. Therefore the first hypothesis has been confirmed in the case of the Skill variable, but not in the case of the Frequency variable. This indicated that the additional efforts to support the learning of Google Drive needed to concentrate on filling any existing gaps in the students' practical skills, without necessarily requiring additional homework requiring repetitive or frequent practice.

TABLE I.
CORRELATIONS FOR GOOGLE DRIVE USAGE VARIABLES

|  | Would recommend GD | Satisfied with GD | Can work effectively with GD | Finds GD fun to use |
|---|---|---|---|---|
| Frequency of using Google Drive (GD) | 0.281 | 0.109 | 0.266 | 0.041 |
| Skill at using GD | 0.635 | 0.406 | 0.543 | 0.553 |

Table 2 summarizes the results of the descriptive statistics for the second hypothesis. The survey questions could be answered according to a five-point Likert scale as follows: strongly disagree (=1), disagree (=2), neutral (=3), agree (=4), and strongly agree (=5). The means of the cumulative responses to the three questions reveal that the students found it helpful to see other group members' work and also found it useful to receive suggestions from other group members who could view their work. The majority of the students answered these two questions with strongly agree. The standard error and standard variance for the responses to these two questions were both small (seeing others and taking suggestions).

TABLE II.
DESCRIPTIVE STATISTICS FOR THE COLLABORATION VARIABLES

|  | Seeing Others | Taking Suggestions | Allowing Co-editing |
|---|---|---|---|
| Mean | 4.67 | 4.71 | 3.38 |
| Median | 5.00 | 5.00 | 3.00 |
| Standard Error | 0.12 | 0.09 | 0.28 |
| Standard Deviation | 0.56 | 0.46 | 1.38 |
| Sample Variance | 0.32 | 0.22 | 1.90 |
| Minimum - Maximum | 3 - 5 | 4 – 5 | 1 - 5 |

However, the students seemed neutral toward the idea of allowing other group members to contribute by editing their individual documents. Furthermore, the standard deviation and the standard error were large, as those student responses varied greatly, including an occurrence of all of the five possible responses all the way from strongly disagree to strongly agree. Therefore the second hypothesis was not confirmed in the case of the first two questions, as students were ready to carry out those two collaborative approaches. However, the second hypothesis was confirmed for the third question, the students were not prepared to collaborate by allowing other students to contribute to their individual documents.

This result also required the lecturer to formulate and refine his future instructions to the students during the course of the project, regarding group editing of certain documents. The following approach was agreed upon: co-editing would not be required in the first phase of the assignment and students would be recommended to limit the sharing to only viewing during that phase. This also addressed any concerns about collaboration possibly affecting individual grades for the first part of the assignment. For the second phase of the assignment, as part of an iterative and integrating process of analysis and design, those project documents would be opened to editing by others, and the assessment for those documents would then become graded as a group in the new phase. Furthermore, the students were recommended, if time allowed, to bring their ideas and suggested changes to the group meetings and discuss them prior to editing the documents online.

The qualitative student responses by the end of Semester 1 of 2014 show that, in comparison with 2013, positive feedback has increased, especially in regards to using Google Drive as a repository for group project deliverables and artefacts. One student echoed the opinions of many of his classmates: "I will be using Google Drive in future assignments." This may be partly due to the lecturer acting upon the results of the diagnostic survey, by promoting the tool with more commitment, and playing a more proactive role by helping students to improve their skills.

Many students have mentioned interpersonal issues of arranging meetings with their team members and using their time efficiently outside of the classroom. One of the ways that many students used Google Drive was to alleviate the problem of not being able to meet in person very often. A student commented: "Google Drive is indispensable. It greatly improved project coordination and sharing of ideas." Another student mentioned how they used Google Hangouts in conjunction with Google Drive. Furthermore, a different but common and effective practice was "using Google Drive during the group meeting." In some groups, all of the students were already Facebook friends with each other. This resulted in an interesting combination of using Facebook in parallel with Google Drive. For example, instant reminders could be sent through Facebook, where they could also share links to Google Documents.

Another outcome that permeated the students' qualitative responses is their own appreciation for what they have accomplished during the project assignment. When group members had a look at and talked about each other's work, these discussions "reduced errors" in their documents prior to submission for grading. Another student commented: "a very good assignment overall feel I learnt a lot". One of the detailed responses described how each member brought unique qualities and "aided in the creation of a final finished assignment that I was proud to be involved in." Not all responses or experiences were positive. Some students reported situations where an individual member did not share certain documents with them. Another erroneous but unusual situation was when certain group members worked on documents offline, and basically removed and re-uploaded those documents to the online folder, instead of consistently maintaining those deliverables online. Overall, the cumulative learner response from this survey have been positive and have provided some reassurance for using the same collaborations tools and practices in future Systems Analysis and Design or other similar courses.

## V. CONCLUSION

Team skills and group learning skills are important in a Systems Analysis and Design course. These attributes are also expected by the industry and useful in the future workplace. Therefore, the design of this particular IT course has paid particular attention to collaboration, and a collaborative technology has been integrated within the delivery of this course. Small group interaction and informal learning are both transformed and taken to a higher level when collaboration and sharing of artefacts takes place online [12]. Furthermore, rapid collaboration is even more important for a systems analysis and design project if the desired product is a mobile or web based application with a limited scope and is expected to be developed in a short period of time. Using online collaboration can help students and professionals during tasks such as prototyping, forward engineering, programming, and testing.

During the last two years, the Systems Analysis and Design online course web site has been updated with more links to open education resources. The current collection of online resources and activities forms a solid foundation; however the SAD blended course can get even better with further open and cloud resources in the future. Relationships with other institutions can also be increased, while utilizing cloud based storage and collaboration tools next to Moodle with links to various Google Documents (e.g. drawings, presentations, tables, and sheets).

Google Drive, as a learning tool, has been implemented in this study with pedagogical goals in mind, and reinforced with the help of student feedback and by providing technical assistance and encouragement. The results of this study are also interesting because these may serve as insights and suggestions for best practices for using Google Drive in the professional workplace. Therefore, the same planning, evaluation, and learner support strategies implemented in this study can also be used in adult and corporate training settings.

In a broader social context, this technology also helps students form closer connections with each other, outside of the classroom. The connections between these individuals will continue after the course, as a part of their personal Google circles and online accounts. Social bonding plays an important role on the Internet by influencing how individuals, e.g. students, tend to behave online [6]. During this study, it was expected that using online collaboration in the course would give students an additional medium for their social needs and interacting with their classmates. However, although social factors encourage the use of internet based technologies, they may also impose some constraints on how an online medium is used, namely in the form of social norms. Individual responsibility is an influential social norm and requires consideration in designing group assignments and group leaning activities. Nevertheless, for the incoming and future young learners of the so-called Generation Z who have grown up with social media, using online collaboration will be an essential component of active learning and teaching strategies.

AUTHOR

**Emre Erturk** is a Senior Lecturer in the School of Computing at the Eastern Institute of Technology, New Zealand. He earned his PhD from the University of Oklahoma in the USA in 2007. Since then, he has also taught face-to-face and distance education with the University of Maryland. His research interests include online and blended education. Email: eerturk@eit.ac.nz.